\documentclass[twocolumn,nofootinbib,preprintnumbers,amsmath,amssymb,superscriptaddress]{revtex4}
\pdfoutput=1

\usepackage[mathletters]{ucs}
\usepackage[utf8]{inputenc}

\usepackage{amsmath,amssymb,amsthm,amsfonts,mathtools}
\usepackage{mathrsfs}
\usepackage{nicefrac}
\usepackage{slashed}
\usepackage{graphicx}
\usepackage{subfigure}
\usepackage[dvipsnames]{xcolor}
\usepackage{rotating}
\usepackage{dsfont}
\usepackage{setspace}
\usepackage{verbatim}
\usepackage{fancyhdr}
\usepackage[pdftitle={Reducing Autocorrelation Times in Lattice	Simulations with Generative Adversarial Networks},
		    pdfauthor={Jan M. Pawlowski, Julian M. Urban}]{hyperref}
\hypersetup{
	colorlinks=true,     
	linkcolor=blue,      
	citecolor=blue,      
	filecolor=blue,      
	urlcolor=blue        
}
\usepackage[nameinlink]{cleveref}
\usepackage{cleveref}
\usepackage{tabularx}

\DeclareMathOperator*{\argmin}{arg\,min}

\begin{document}

\title{Reducing Autocorrelation Times in Lattice Simulations\\ with Generative Adversarial Networks}

\author{Jan M.~Pawlowski}
\affiliation{Institut f\"ur Theoretische
	Physik, Universit\"at Heidelberg, Philosophenweg 16, 69120
	Heidelberg, Germany}
\affiliation{ExtreMe Matter Institute EMMI, GSI,
	Planckstra{\ss}e 1, 64291 Darmstadt, Germany}

\author{Julian M. Urban}
\affiliation{Institut f\"ur Theoretische
	Physik, Universit\"at Heidelberg, Philosophenweg 16, 69120
	Heidelberg, Germany}

\begin{abstract}
Short autocorrelation times are essential for a reliable error
assessment in Monte Carlo simulations of lattice systems. In many
interesting scenarios, the decay of autocorrelations in the Markov
chain is prohibitively slow. Generative samplers can provide
statistically independent field configurations, thereby potentially
ameliorating these issues. In this work, the applicability of neural
samplers to this problem is investigated. Specifically, we work with a
generative adversarial network (GAN). We propose to address
difficulties regarding its statistical exactness through the
implementation of an overrelaxation step, by searching the latent
space of the trained generator network. This procedure can be
incorporated into a standard Monte Carlo algorithm, which then permits
a sensible assessment of ergodicity and balance based on consistency
checks. Numerical results for real, scalar $\phi^4$-theory in two
dimensions are presented. We achieve a significant reduction of
autocorrelations while accurately reproducing the correct statistics.
We discuss possible improvements to the approach as well as potential
solutions to persisting issues.
\end{abstract}
\maketitle

\section{Introduction}
\label{sec:intro}

Minimizing the statistical error is essential for applications of
lattice simulations. A reliable assessment of this error requires
sufficiently short autocorrelation times in the Markov chain. This
becomes a problem in models affected by critical slowing down, which
is a severe hindrance for the extrapolation of lattice calculations to
the continuum. This issue has inspired a decades-long search for ever
more efficient sampling techniques. Simulations of this kind have been
and continue to be vital for our understanding of fundamental
interactions from first principles.

More recently, promising new approaches based on generative machine
learning methods are being explored. An interesting candidate for this
purpose is the generative adversarial network (GAN) \cite{1406.2661},
which has lately received much attention in the machine learning
community. Its potential for lattice simulations has been demonstrated
for the two-dimensional Ising model \cite{1710.04987} and complex
scalar $\phi^4$-theory at non-zero chemical potential
\cite{1810.12879}. By construction, the samples are generated
independently. Hence, in principle, there are no autocorrelations if
they are sequentially arranged in a Markov chain. This makes GANs
highly attractive in the quest for more efficient simulation
algorithms. Other approaches studied in this context are the
restricted Boltzmann machine \cite{PhysRevB.95.035105,
	Tanaka:2017niz}, flow-based generative models \cite{Albergo:2019eim,
	kanwar2020equivariant}, autoregressive networks
\cite{PhysRevLett.122.080602, Nicoli2019asymptotically,
	Nicoli2019comment} and the self-learning Monte Carlo method
\cite{PhysRevB.95.041101, PhysRevB.95.241104, PhysRevB.97.205140,
	PhysRevB.98.041102, 1807.04955}. Machine learning methods in general
are also increasingly used for a variety of other tasks in high energy
physics and condensed matter theory \cite{PhysRevB.94.195105,
	science.aag2302, nphys4035, s41598-017-09098-0, PhysRevX.7.031038,
	nphys4037, j.jcp.2017.06.045, PhysRevX.7.021021, s41467-017-00705-2,
	s41567-018-0048-5, PhysRevE.95.062122, PhysRevB.96.205146,
	Morningstar:2017:DLI:3122009.3242020, s0217984916504017,
	PhysRevB.94.165134, Cristoforetti:2017naf, PhysRevB.96.205152,
	PhysRevD.97.094506, Funai:2018esm, PhysRevE.96.022140,
	PhysRevB.96.184410, Kades2019, bluecher2020novel,
	wang2020recognizing}, see also \cite{Mehta_2019} for an introduction
and \cite{carrasquilla2020machine} for a recent review.

However, simply replacing a Monte Carlo algorithm with a GAN is
problematic from a conceptual point of view. The learned distribution
typically shows non-negligible deviations from the desired target.
This casts some doubt on the reliability of such an approach.
Moreover, even if the approximation by the network exhibits high
precision, one cannot simply assume that sampling from it is
sufficiently ergodic. Implementing these key properties in a
conclusive manner is essential for accurate and reliable lattice
computations.

In this paper, we propose the implementation of an overrelaxation step
using a GAN, in combination with a traditional hybrid Monte Carlo
(HMC) algorithm. This approach effectively breaks the Markov chain for
observables unrelated to the action, thereby leading to a reduction in
the associated autocorrelation times. We put forward self-consistency
checks and numerical arguments to assess whether the sampling is both
sufficiently ergodic and statistically exact, and thus correctly
captures the dynamics of the theory. Conditions for asymptotic
exactness as well as potential efficiency gains are discussed
extensively.

The approach is demonstrated in the context of real, scalar
$\phi^4$-theory in two dimensions. First, we show that a simple
`vanilla' GAN can reproduce observables in the disordered phase of the
theory with high accuracy from a comparatively small number of
training samples. We then demonstrate that by introducing the GAN
overrelaxation step, a significant reduction of the autocorrelation
time of the magnetization can be achieved (\Cref{fig:autocorr}). We
argue that such an approach could greatly improve the computational
efficiency of traditional sampling techniques and our results motivate
further research into the matter.

The paper is organized as follows. In \Cref{sec:phifourth} we briefly
review $\phi^4$-theory on the lattice. \Cref{sec:metropolis} recalls
relevant aspects of the Metropolis-Hastings algorithm as well as the
overrelaxation method. \Cref{sec:gans} serves to introduce GANs and
illustrate their potential for improved sampling algorithms. Our
approach is developed in \Cref{sec:algorithm}. Numerical results are
presented in \Cref{sec:results}. We summarize our findings and sketch
strategies for future research in \Cref{sec:concl}.

\section{Scalar $\phi^4$-Theory on the Lattice}
\label{sec:phifourth}

We work with real, scalar $\phi^4$-theory discretized on a
two-dimensional Euclidean square lattice with periodic boundary
conditions. We consider only isotropic, symmetric lattices spanning
$N$ sites in each dimension. The associated action in its
dimensionless form is given by
\begin{align}\nonumber 
  S = &\,\sum\limits_{x \in \Lambda} \Biggl[-2\kappa \sum\limits_{\mu=1}^d
        \phi(x) \phi(x+\hat{\mu}) \\
  &\hspace{2cm}+ (1 - 2\lambda) \phi(x)^2 + \lambda \phi(x)^4 \Biggr]\ .
  \label{action}
\end{align}
Here, $\Lambda$ denotes the set of all lattice sites and $\hat{\mu}$
the unit vector in $\mu$-direction. $\kappa$ is commonly called the
Hopping parameter and $\lambda$ parametrizes the coupling strength of
the quartic interaction.

The theory belongs to the Ising universality class. As such, in $d =
2$ it exhibits a phase transition associated with spontaneous breaking
of the $Z_2$ symmetry. The order parameter is the magnetization,
defined as the expectation value of the field,
\begin{align}
  \langle M \rangle = \left\langle \frac{1}{V}
  \sum_{x \in \Lambda} \phi(x) \right\rangle\, ,
\end{align}
where $V = |\Lambda| = N^d$ denotes the dimensionless volume, i.e.\
the number of lattice sites. $\langle M \rangle$ assumes a non-zero
value in the broken phase. In the thermodynamic limit,
this phase transition is characterized by a divergence in the
connected two-point susceptibility,
\begin{align}	\label{chi2}
	\chi_2 = V \left(\langle M^2 \rangle - \langle M \rangle^2 \right)\ .
\end{align}
On a finite-size lattice, one instead observes a peak, which then
narrows as the volume is increased. We also consider the Binder
cumulant
\begin{align}
	U_L = 1 - \frac{1}{3} \frac{\langle M^4 \rangle}{\langle M^2 \rangle^2} \ ,
\end{align}
which quantifies the curtosis of the fluctuations.

\section{Metropolis-Hastings, Overrelaxation and Critical Slowing Down}
\label{sec:metropolis}

The `gold standard' of Monte Carlo sampling methods in statistical
physics is the Metropolis-Hastings algorithm \cite{Metropolis:1953am}.
It explores the configuration space through aperiodic updates ensuring
ergodicity, combined with an accept/reject rule to asymptotically
ensure statistical exactness in the limit of large numbers of samples.
Suppose we wish to sample field configurations from the Boltzmann
weight $P(\phi)\ \propto\ \exp (-S[\phi])$. A candidate $\phi'$ is
generated from the current configuration $\phi$ with some a priori
selection probability $T_0(\phi'|\phi)$. It is then accepted or
rejected according to the acceptance probability
\begin{align}\label{eq:acceptance1}
          T_A(\phi'|\phi) = \min \left(1,\ \frac{T_0(\phi|\phi')
              \exp(-S[\phi'])}{T_0(\phi'|\phi) \exp(-S[\phi])}\right) \ .
\end{align}
One usually considers a symmetric selection probability
$T_0(\phi|\phi') = T_0(\phi'|\phi)$, the special case commonly being
referred to as just the Metropolis algorithm. \Cref{eq:acceptance1}
then simplifies to
\begin{align}\label{eq:acceptance2}
	T_A(\phi'|\phi) = \min(1,\ \exp(-\varDelta S)) \, ,
\end{align}
where $\varDelta S = S[\phi'] - S[\phi]$. For a real, scalar field,
this is ensured by proposing updates $\phi'(x)$ distributed
symmetrically around $\phi(x)$, such that $\phi'(x) - \phi(x)$ is zero
on average.

Local updating methods based on the Metropolis-Hastings algorithm
usually exhibit long autocorrelation times. Significant improvements
can be achieved with the HMC method. It is based on a molecular
dynamics evolution using classical Hamiltonian equations of motion,
combined with an accept/reject step. This allows larger steps with
reasonable acceptance rates.

A technique used to speed up the motion through configuration space is
overrelaxation. It was originally designed for simulations of $SU(2)$
and $SU(3)$ Yang-Mills theory by exploiting symmetries of the action
\cite{PhysRevD.36.515}. It is based on the fact that a candidate
configuration is automatically accepted in the Metropolis step if
$\varDelta S = 0$, under the condition that $T_0$ is symmetric. This
can be achieved by performing rotations in group space that leave $S$
unchanged. By itself, overrelaxation is therefore not ergodic, since
it moves on the subspace of constant action. Ergodicity is achieved by
combining it with a standard Monte Carlo algorithm.

\begin{figure*}[t]
	\centering
	\makebox[\textwidth][c]{\includegraphics[width=0.85\linewidth]{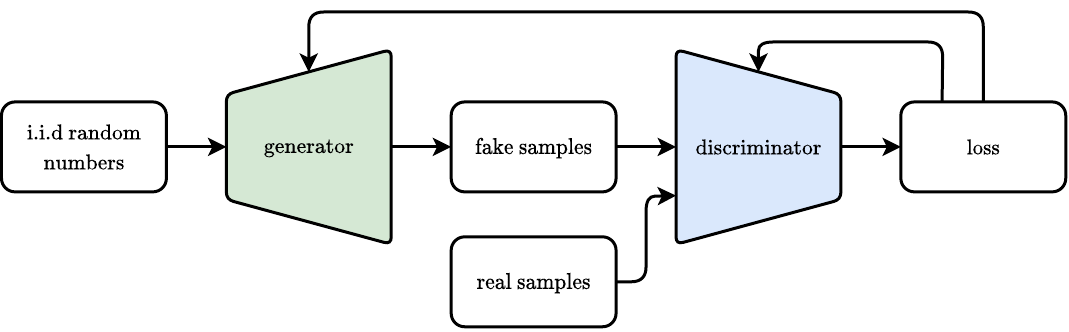}}
	\caption{Simple schematic of a GAN's components and data flow. I.i.d.\
	random numbers are passed to the generator to produce fake samples.
	The discriminator learns to distinguish between real and fake.
	Training is performed by backpropagating loss gradients through both
	networks and updating weights in an alternating fashion.}
	\label{fig:gan}
\end{figure*}

\begin{figure*}[t]
	\centering
	\makebox[\textwidth][c]{\includegraphics[width=0.85\linewidth]{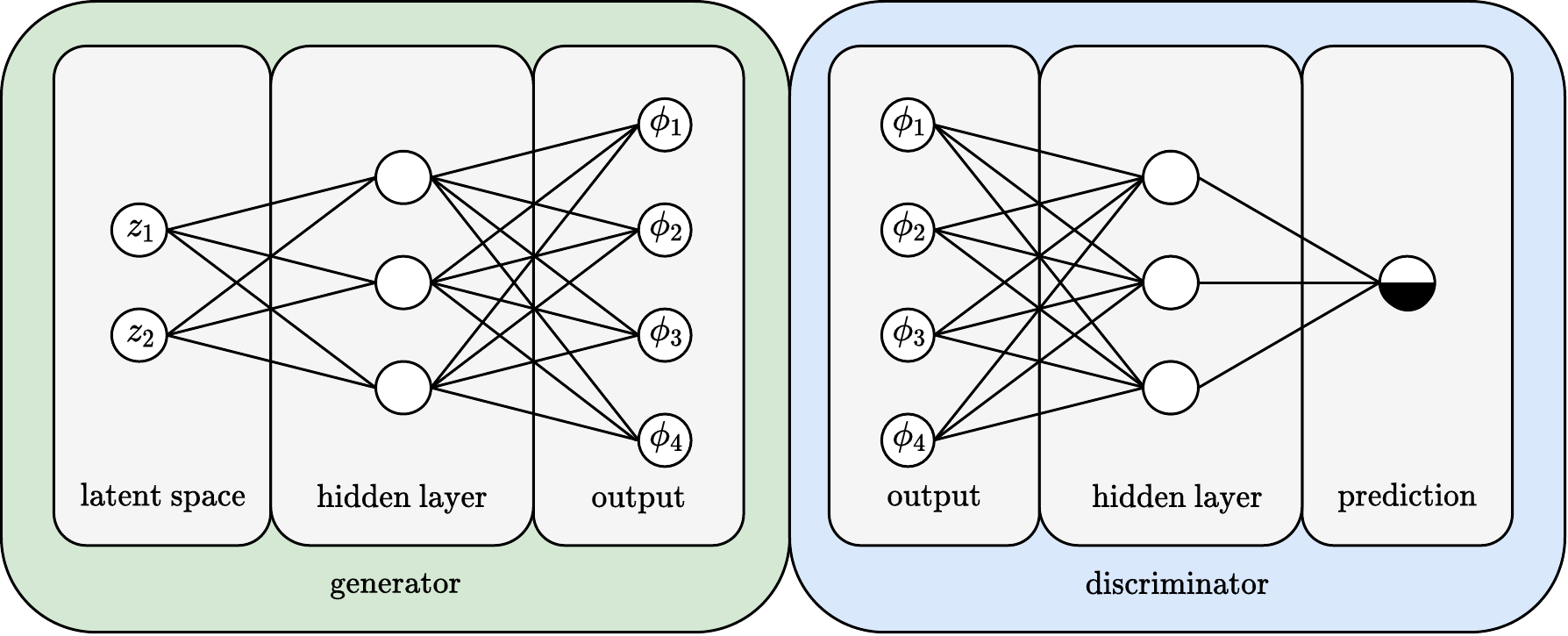}}
	\caption{Illustration of the vanilla GAN architecture with
	fully-connected layers. Neurons are depicted as circles and synapses
	as lines. The output layer of the discriminator features a single,
	binary neuron whose state determines the prediction as real or fake.}
	\label{fig:gan_structure}
\end{figure*}

However, neither the HMC algorithm nor the overrelaxation method solve
critical slowing down. Simply put, the problem occurs whenever the
correlation length $\xi$ of a system diverges. Define the
autocorrelation function of an observable $X$ as
\begin{align}
	\begin{split}
	C_X(t) &= \langle (X_i - \langle X_i \rangle )\ (X_{i+t} - \langle X_{i+t} \rangle ) \rangle\\
	&= \langle X_i X_{i+t} \rangle - \langle X_i \rangle \langle X_{i+t} \rangle \ ,
	\end{split}
\end{align}
where $i$ is the position and $t$ denotes the discrete time of the
Markov chain, i.e.\ the number of steps. Typically, $C_X(t)$ decays
exponentially,
\begin{align}
	C_X(t) \sim \exp \left(- \frac{t}{\tau_{\text{exp}}}\right)\ .
\end{align}
Here, $\tau_{\text{exp}}$ denotes the exponential autocorrelation
time, which one expects to scale as a power of the correlation length,
$\tau_{\text{exp}} \sim \xi^z$. The dynamical critical exponent $z
\geq 0$ depends on the type of algorithm used. Close to a critical
point, the correlation length diverges. This hampers the investigation
of critical phenomena as well as extrapolations to the continuum
limit.

\section{Generative Adversarial Networks}
\label{sec:gans}

The GAN belongs to a class of unsupervised, generative machine
learning methods based on deep neural networks. The characteristic
feature that distinguishes it from many other architectures is the
utilization of game theory principles for its training. It consists of
two consecutive feedforward neural networks (see \Cref{fig:gan} for a
schematic), the generator $G$ and discriminator $D$. They are
non-linear, differentiable functions whose learnable parameters are
commonly called weights. The $d_z$-dimensional input vector $z$ of the
generator is drawn from a multi-variate prior distribution $P(z)$. The
discriminator receives the generator outputs $G(z)$ as well as samples
$\phi$ from the training dataset. $D$ is a classification network with
binary output that is trained to distinguish between `real' and `fake'
samples. Its last layer consists of a single neuron with a sigmoid
activation, which allows to define a loss function in terms of the
binary cross-entropy. Choosing the labels for real and fake samples
arbitrarily (but consistently) as 0 or 1, the loss per prediction with
label $y$ is defined as

\begin{align}\label{eq:crossentropy}
	L_\text{BCE} = -y \log &(D(G(z))) \nonumber \\&+ (y - 1) \log (1 - D(G(z)))\, .
\end{align}
Training corresponds to minimizing the loss separately for $D$ and $G$
using opposite labels, respectively. This is achieved by evolving
their weights according to a gradient flow equation in an alternating
fashion. The gradient of the loss with respect to the weights is
calculated using automatic differentiation, and then backpropagated
through both networks by the chain rule. In intuitive terms, the
optimization objective for the discriminator is to maximize its
accuracy for the correct classification of $\phi$ and $G(z)$ as real
or fake. Simultaneously, the generator is trained to produce samples
that cause false positive predictions of the discriminator, thereby
approximating the true target distribution $P(\phi)$. The two networks
play a zero-sum non-cooperative game, and the model is said to
converge when they reach so-called Nash equilibrium.

This entails the fundamental difference and possible advantage of this
method compared to traditional Monte Carlo sampling. Since $P(z)$ is
commonly chosen to be a simple multi-variate uniform or Gaussian
distribution, candidate configurations drawn from a properly
equilibrated generator are by construction statistically independent.
This is because the $z$ are sampled i.i.d.\ and the resulting field
configurations are not calculated as consecutive elements of a
traditional Markov chain.

In practice however, one encounters deviations of varying severity
from the true target distribution. Also, GANs may not be sufficiently
ergodic in order to perform reliable calculations. This becomes
apparent when one considers the extreme case of so-called mode
collapse, where the generator learns to produce only one or a very
small number of samples largely independent of its prior distribution.
Insufficient variation among the GAN output is not punished by the
discriminator and can only be checked a posteriori. A number of
improved approaches to deal with such issues have since been proposed
in the literature \cite{gui2020review}. Still, one may question
whether GANs can be sufficiently random for the level of rigor
required in certain statistical sampling problems. Interestingly, it
was shown that they can act as reliable pseudo-random number
generators, outperforming several standard, non-cryptographic
algorithms \cite{bernardi2018pseudorandom}.
\begin{figure}[b]
	\includegraphics[width=\linewidth]{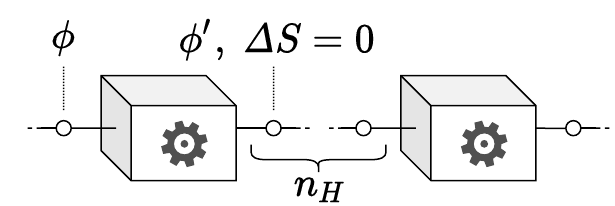}
	\caption{Sketch of the updating procedure on the Markov chain, with
	boxes depicting the overrelaxation step.}
	\label{fig:overrelaxation}
\end{figure}

\section{Algorithmic Framework}
\label{sec:algorithm}

Using GANs to accelerate lattice computations in a reliable manner
requires implementing a statistical selection procedure as well as
defining sensible criteria to assess the validity of such an approach.
Naively, one could just try to equip the GAN with a Metropolis
accept/reject step, thereby simply hoping that it is sufficiently
ergodic. However, even with all statistical concerns aside, this is
not feasible in practice, for the simple reason that candidate
configurations are accepted either automatically if $\varDelta S \leq
0$ or with probability $\exp(-\varDelta S)$ if $\varDelta S > 0$.
Accordingly, such an algorithm only works if changes in the action are
not too large. Common Monte Carlo algorithms can usually be tuned to
avoid this problem and achieve reasonable acceptance rates. For the
GAN, large positive and negative values for $\varDelta S$ would be
common, since subsequent samples are uncorrelated. Hence, the
algorithm effectively freezes at the lower end of the available action
distribution after a short time. Jumping to larger values of $S$ is in
principle possible, but exponentially suppressed, leading to a
vanishing acceptance rate. This thought experiment emphasizes the need
for an asymptotically unbiased statistical selection procedure and
convincing numerical tests thereof.

\subsection{Overrelaxation with GANs}

In order to avoid the aforementioned issues, we propose to implement
the GAN as an overrelaxation step, which can then be integrated into
any action-based importance sampling algorithm. In this manner,
autocorrelations can be reduced substantially while still
asymptotically approaching the correct distribution of the theory. Our
method of selecting suitable candidate configurations is based on
\cite{1805.07281}, where GANs are proposed as an ansatz to the more
general task of solving inverse problems.

Our approach is implemented with the following procedure (see
\Cref{fig:overrelaxation} for a sketch):
\begin{itemize}
	\item Take a number of HMC steps $n_H$ to obtain a configuration
$\phi$. In this work, only accepted samples are counted by $n_H$ to
facilitate a consistent number of HMC trajectories between GAN
overrelaxation steps in the Markov chain.
	\item Pre-sampling: Sample from the GAN until a configuration $G(z)$
is found that fulfills $|\varDelta S| = |S[G(z)] - S[\phi]| \leq
\varDelta S_\text{thresh}$. $\varDelta S_\text{thresh}$ is tuned
such that $S[\phi]$ and $S[G(z)]$ are close, which accelerates the
gradient flow step, but also such that a suitable $G(z)$ can be
generated in a reasonable amount of time.
	\item Gradient flow: Perform a gradient descent evolution of the
associated latent variable $z$ using $\varDelta S^2$ as loss
function, i.e.
	\begin{align}
		z' = \argmin_z (S[G(z)] -  S[\phi])^2\ ,
	\end{align}
	
	by employing a standard discretized gradient flow equation,
	
	\begin{align}\label{eq:gradient}
		z'(\tau + \epsilon) = z'(\tau) - \epsilon \frac{\partial \varDelta S^2}{\partial z'}\ .
	\end{align}
	
	As mentioned in the last section, gradient flow is also commonly
applied to learn the optimal weights of a network. In this context,
the finite step size $\epsilon$ is called learning rate, and
gradients are usually obtained by automatic differentiation. For this
work, we use the Adam algorithm \cite{Adam}, a particular variant of
stochastic gradient descent.
\end{itemize}

In this manner, $S[\phi]$ and $S[G(z')]$ can be matched arbitrarily
well, down to the available floating point precision. The action
values can then be considered effectively equal for all intents and
purposes. In principle, this evolution can be performed for any
randomly drawn $z$ without the need for repeated sampling until a
configuration with $|\varDelta S| < \varDelta S_\text{thresh}$ is
found. The additional pre-sampling step simply ensures that the
distance in the latent space between the initial value for $z$ and the
target $z'$ is already small a priori, which speeds up the evolution
and avoids the risk of getting stuck in a local minimum of the loss
landscape. The specific choice of $\varDelta S_\text{thresh}$,
$\epsilon$ as well as the sampling batch size should be determined
through a hyperparameter optimization in order to maximize the
efficiency.

\subsection{Statistical Properties and Consistency Checks}

After the evolution is complete, $G(z')$ is proposed as the new
candidate configuration $\phi'$. Under rather mild assumptions, we
expect the GAN's selection probability $T_0(\phi'|\phi)$ to be
symmetric, as required for the simplification of \Cref{eq:acceptance1}
to (\ref{eq:acceptance2}). We now provide a heuristic
argument for this statement, but also emphasize that this essential
property should always be verified by additional numerical tests,
which we detail below.

Since every field configuration $\phi$ corresponds to a point
$z(\phi)$ in the latent space\footnote{The existence of points in the
	latent space for every possible field configuration is used here as a
	formal argument, although it may not always be possible to actually
	identify the $z$ corresponding to a specific $\phi$. However, this is
	neither necessary nor desired and does not pose a problem in practice.
	Much like in a standard Monte Carlo simulation, for a successive
	update of $\phi'$ it is effectively impossible that it happens to
	revert to its previous value $\phi$, since all variables and
	randomized update proposals are real numbers with floating point
	precision.}, the selection probability is a priori just that of the
latent variables, $T_0(z(\phi')|z(\phi))$. However, $z(\phi')$ does
not explicitly depend on $z(\phi)$, and the latter is also not
accessible in general. Instead, it only depends on the value of
$S[\phi]$, i.e.\ $z(\phi')$ follows the conditional distribution
$P(z|S[G(z)]=S[\phi])$. Hence, to fulfill the symmetry requirement, we
only need to verify that the distribution of latent variables at a
fixed value of the action is also Gaussian like the full prior $P(z)$.
Random variables drawn from such a multi-variate normal distribution
can of course be sampled i.i.d.\ from uniformly distributed
pseudo-random numbers using the Box-Muller transform. In this case,
the associated selection probability is in fact independent and
exactly 1, $T_0(z'|z) \equiv T_0(z') = 1$. It is precisely at this
point that the Markov chain is broken and autocorrelations are
eliminated for observables unrelated to $S$.

\begin{figure*}[t]
	\centering
	\includegraphics[width=0.49\linewidth]{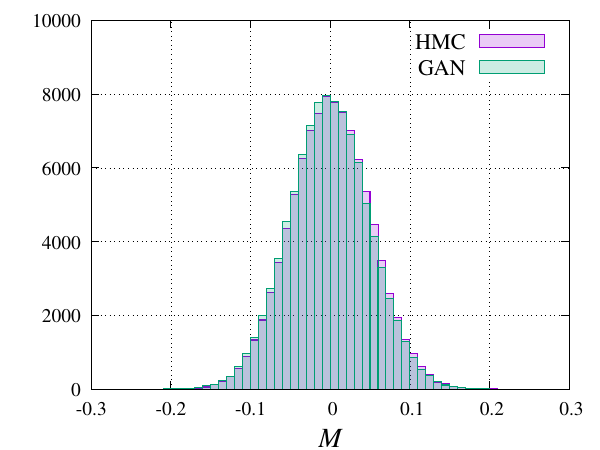}
	\includegraphics[width=0.49\linewidth]{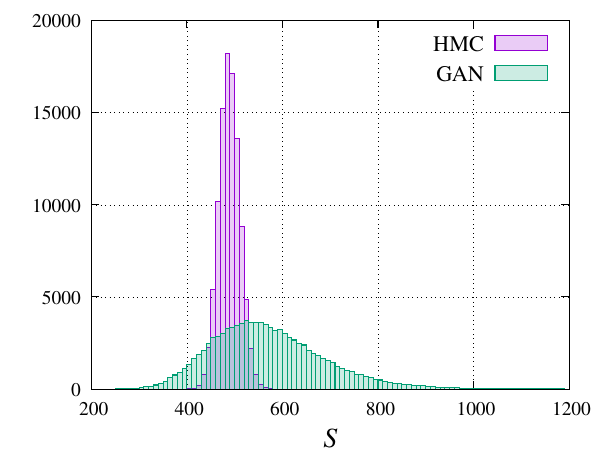}
	\caption{Comparison of magnetization (left) and action (right)
	distributions with $10^5$ samples generated with the HMC algorithm
	and the GAN trained on $10^3$ configurations.}
	\label{fig:gan_baseline}
\end{figure*}

The important observation here is that, in principle, any $z$ can be
chosen as the starting point for the gradient evolution according to
\Cref{eq:gradient}. However, as mentioned above, the argument rests on
the assumption that ensembles of latent variables constrained to the
hypersurfaces of the respective constant actions are also normally
distributed in the latent space. In intuitive terms, we need to check
that the gradient flow evolution that pushes $z$ onto the hypersurface
does not significantly change its distribution $P(z)$. This can be
investigated numerically by estimating average distances between the
initial values of $z$ and $z'$ and comparing raw moments of their
distributions. The changes $\phi'(x) - \phi(x)$ for individual field
variables should then be symmetric on average, which can also be
checked straightforwardly. Since the action remains unchanged by the
whole procedure, the total acceptance probability is therefore also
exactly 1, and the sample is accepted automatically. Hence, under
these conditions, we conclude that the proposed method exhibits the
same statistical properties as traditional overrelaxation.

We now discuss the question of ergodicity. To this end, we first note
the following: for a real scalar field $\phi$ on a $d$-dimensional
lattice, the available configuration space is in principle
$\mathbb{R}^{N^d}$. If we use a multi-variate Gaussian as the prior
distribution $P(z)$, then correspondingly the available latent space
is $\mathbb{R}^{d_z}$. If $d_z < N^d$, this latent space has measure
zero in the target space and the question arises whether sampling from
such an object can even be ergodic. However, simply choosing a large
enough $d_z$ is by no means an easy solution guaranteeing ergodicity,
especially when issues like mode collapse are considered. Then again,
no algorithm is ergodic in a strict sense due to the eventual
periodicity of the underlying pseudo-random number generators,
regardless of how extremely long their periods may be. It is clear
that in order to conduct reliable calculations, the question one needs
to consider is not whether an algorithm is truly ergodic, but
\textit{sufficiently} ergodic. While mathematically it is certainly
possible to construct bijective mappings between $\mathbb{R}^{d_z}$
and $\mathbb{R}^{N^d}$ for any $d_z, N^d \in \mathbb{N}$, these
mappings are not necessarily structure-preserving in any sense.
Considering finite numerical precision, the number of different latent
variables that can possibly be stored in memory, using a specified
number of bits, is in fact smaller than the number of possible field
configurations if $d_z < N^d$. While this may seem problematic from a
conceptual point of view, in practice it is irrelevant, since both
numbers are astronomically large compared to the actual number of
samples commonly needed to accurately determine expectation values of
observables. However, this further highlights the importance of
consistency checks to ensure that the theory is correctly captured by
the network and the number of d.o.f.\ in the latent space is
sufficient.

Based on the discussion above, one can expect that GANs with very
small $d_z$ fail to approximate the target distribution. Above a
certain threshold value one should then observe a plateau where the
performance stabilizes and becomes independent of $d_z$. This
indicates that the size of the latent space has become sufficient for
the GAN to capture the theory's dynamics to a satisfying degree. It is
therefore sensible to train and compare several GANs with different
input dimensions. One should check the network's ability to reach the
whole range of potential action states from its prior distribution. To
this end, it has to be verified that it is possible to generate
matching samples $\phi'$ for every field configuration $\phi$ provided
by a Monte Carlo method such that $\varDelta S = 0$. This
proof-of-work is a necessary criterion as well as a strong indicator
that the algorithm can be sufficiently ergodic, since it shows that
all potential values of the action are in principle accessible through
the GAN's latent space. A first test can simply be performed on a
dataset of independent configurations that have not been used for
training. If the GAN has experienced mode collapse or has not captured
the dynamics of the theory well enough, the overrelaxation step will
not work for the majority of samples. Since in our algorithm the space
of action values is traced out by the ergodic HMC updates,
successfully running the simulation with this method is by itself a
strong argument for sufficient ergodicity.

This concludes the discussion of the proposed algorithmic framework.
The key improvement brought on by our method is that it can
effectively break the Markov chain for observables unrelated to the
action, while preserving essential statistical properties. This leads
to a maximal decorrelation with each GAN overrelaxation step. By
introducing our statistical selection procedure, it is also not
necessary for the network to approximate the target distribution to an
exceedingly high precision. Another important advantage is the
applicability to a much wider variety of models as compared to
conventional overrelaxation, since our algorithm does not depend on
specific symmetries of the action. In fact, for scalar
$\phi^4$-theory, which exhibits only $Z_2$ reflection symmetry, no
such method currently exists.

\section{Numerical Results}
\label{sec:results}

\subsection{Training Details}

To put our algorithm to the test, we first trained a GAN on field
configurations from scalar $\phi^4$-theory on a $32 \times 32$ lattice
as defined in \Cref{sec:phifourth}, using 1000 samples generated in
the symmetric phase at $\kappa = 0.21$ and $\lambda = 0.022$. In order
to provide an estimate of the position in the phase diagram
corresponding to this specific choice of couplings, we note that the
transition occurs at roughly $\kappa \approx 0.27$ (with fixed
$\lambda = 0.022$) \cite{Pawlowski:2017rhn}. We employed one hidden
layer of size 512 for both the generator and the discriminator (see
\Cref{fig:gan_structure} for a schematic). The generator's last layer
has no activation function, thereby allowing values $G(z) \in
\mathbb{R}^{N^d}$. The discriminator's output neuron features a
sigmoid activation, allowing to train for binary classification using
the cross-entropy loss (\Cref{eq:crossentropy}). For all other layers
we used the ReLU activation. The time required for the training was
generally negligible, being of the order of at most a few minutes
until equilibration, in contrast to several days of continuous
sampling on the same hardware to compute the autocorrelation times
discussed below.

\begin{figure}[t]
	\begin{tabular}{c c}
		\includegraphics[width=0.4\linewidth]{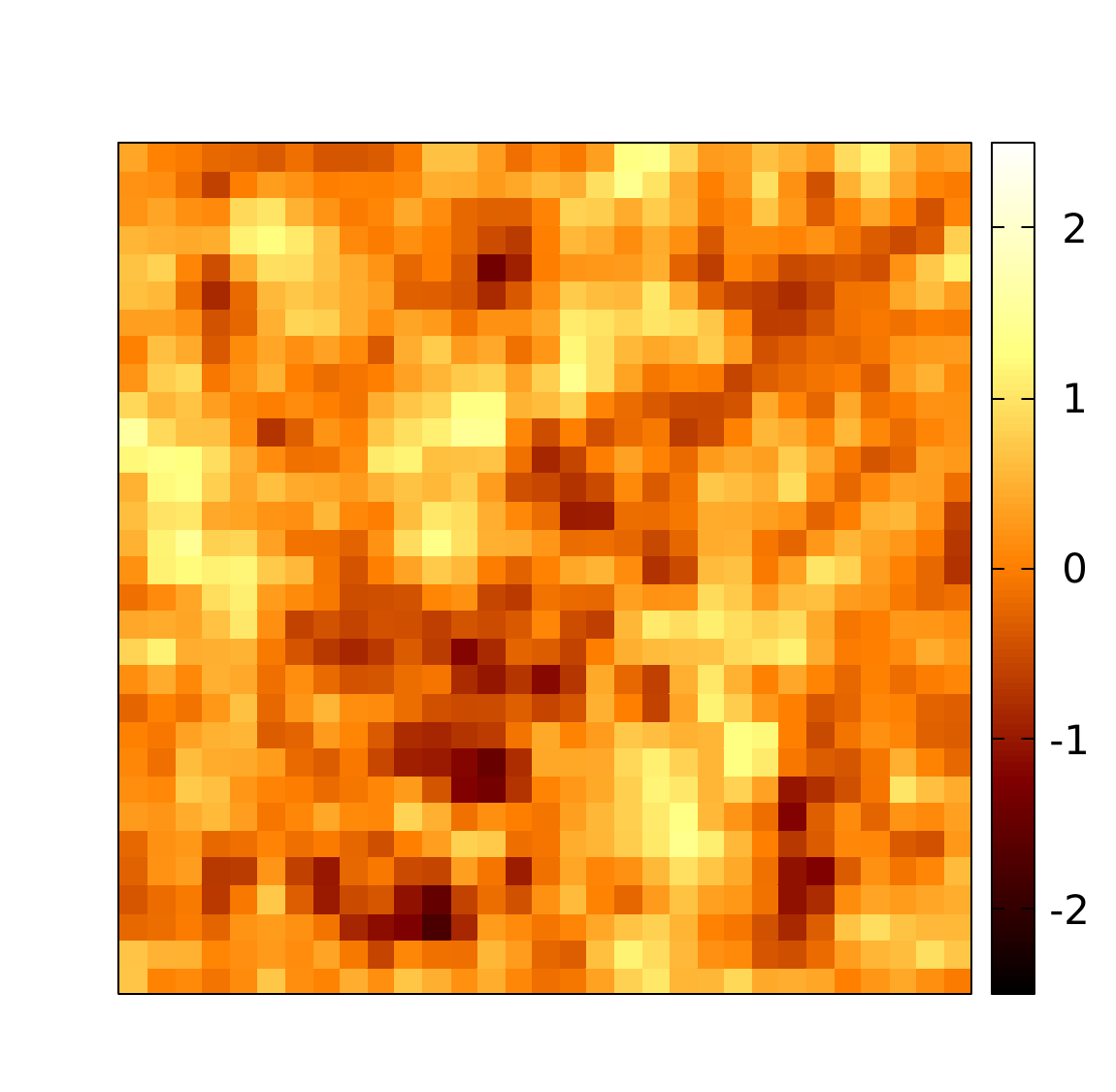} &
		\includegraphics[width=0.4\linewidth]{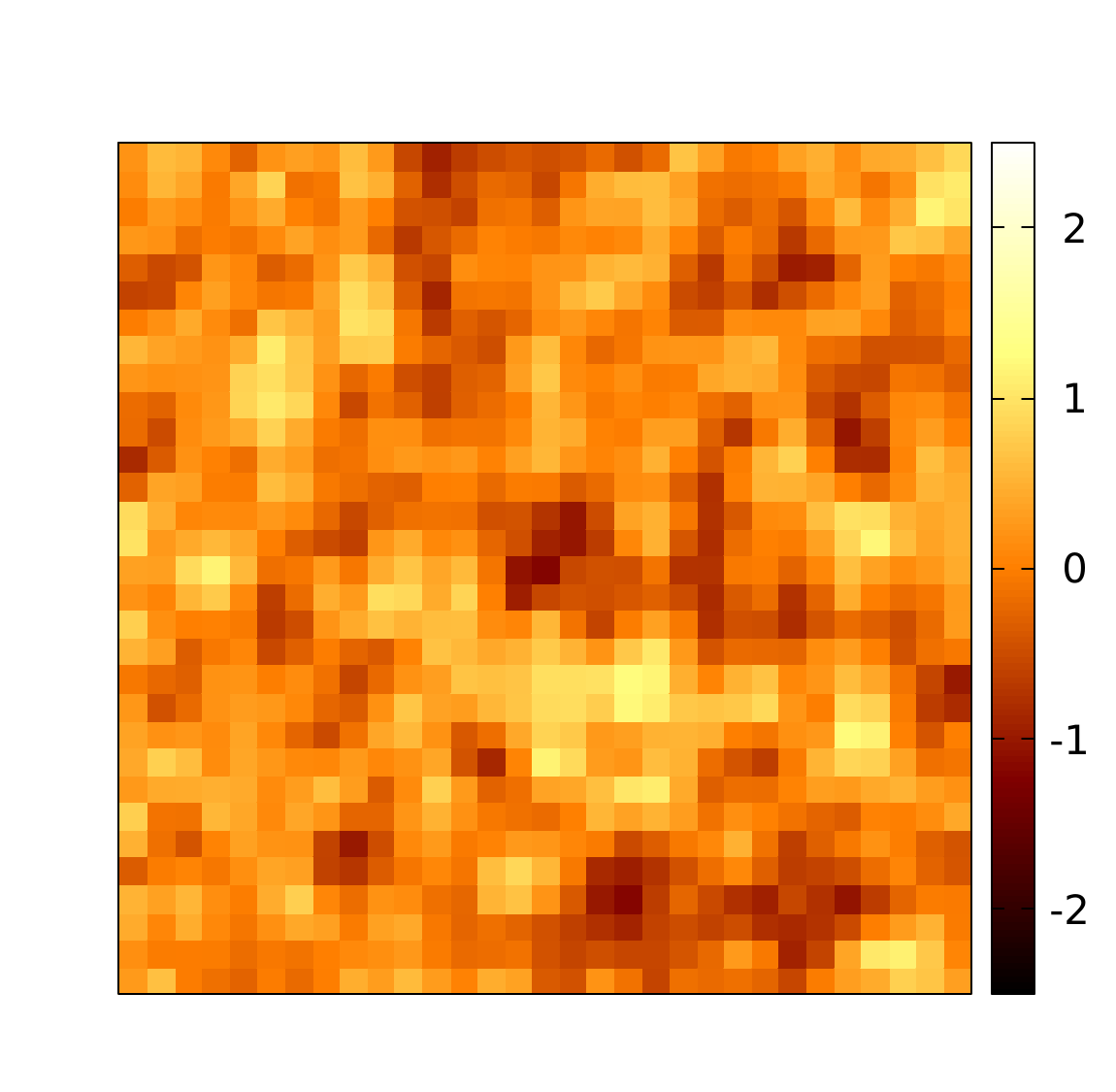}\\
		(a) & (b) \\
		\includegraphics[width=0.4\linewidth]{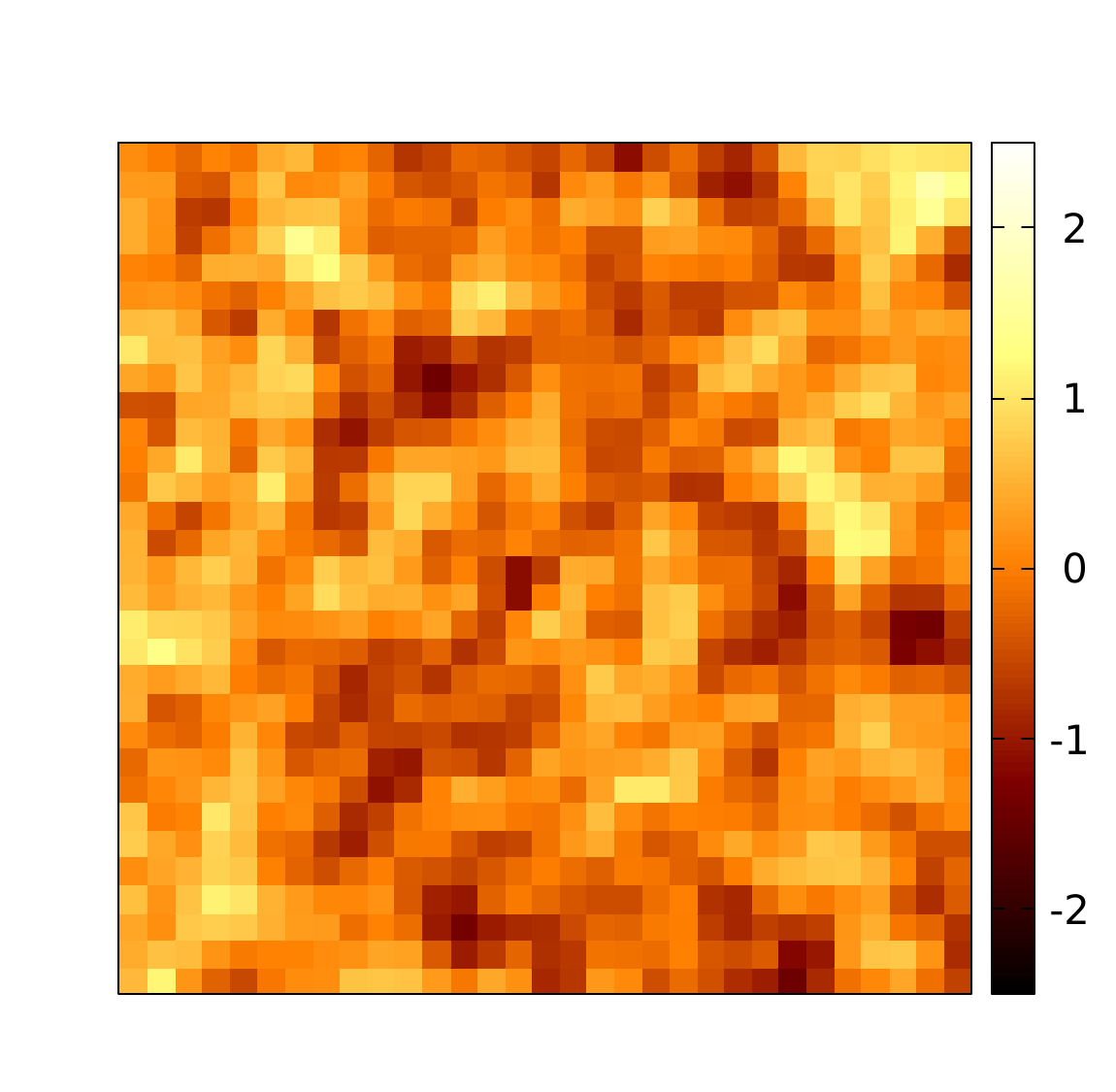} &
		\includegraphics[width=0.4\linewidth]{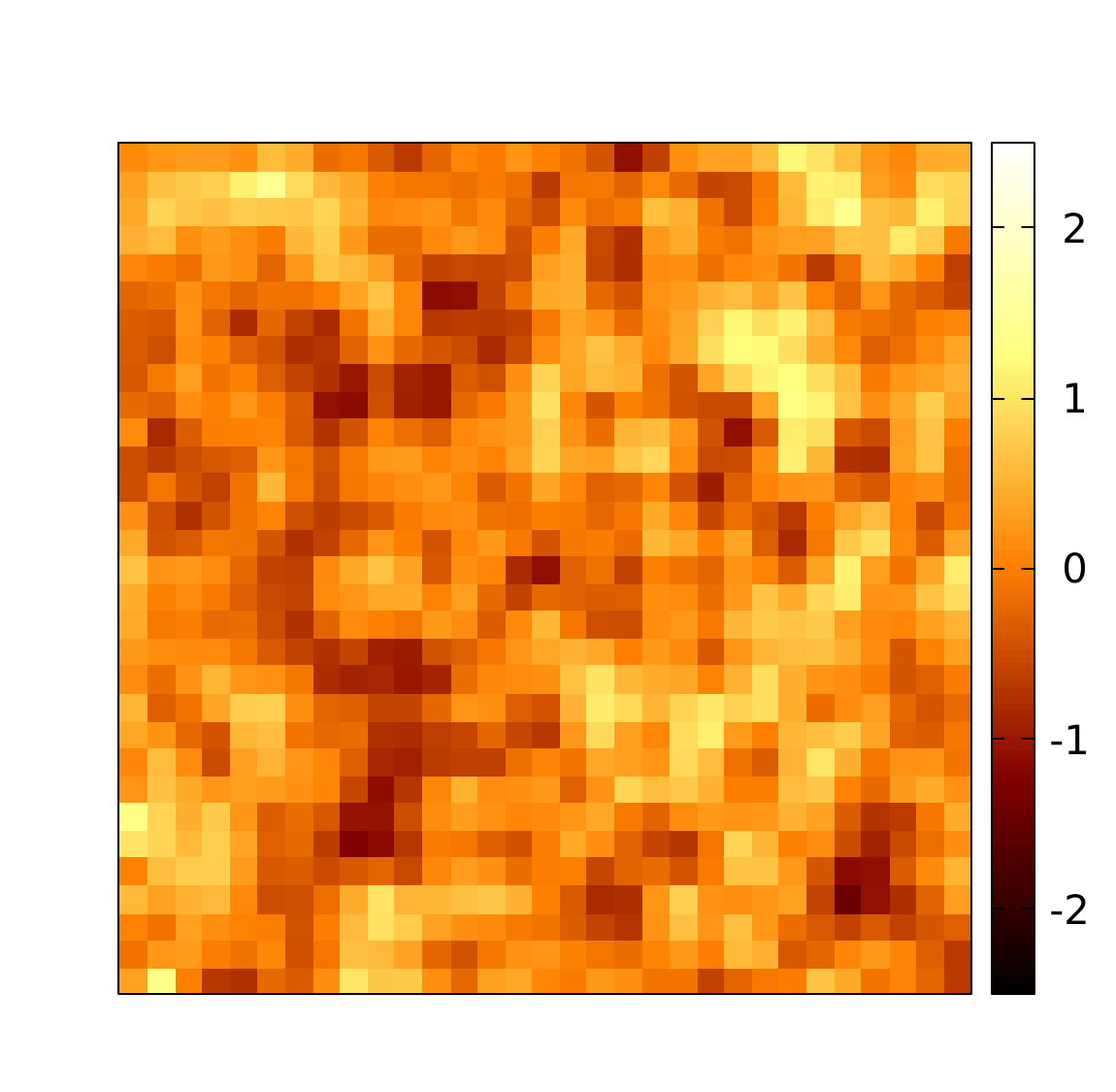}\\
		(c) & (d) \\
	\end{tabular}
	\caption{(a) Field configuration sampled with HMC. (b-d) Samples with
	the same value of the action as (a), generated with GAN
	overrelaxation.}
	\label{fig:samples}
\end{figure}

The distributions of $M$ and $S$ computed with configurations sampled
independently from the GAN are compared to the HMC baseline in
\Cref{fig:gan_baseline}. They already match well for $M$, and several
observables of the theory are reproduced by the GAN to a higher
precision than could be inferred from just the training data.
\Cref{tab:observables} illustrates this for the magnetization $M$, the
two-point susceptibility $\chi_2$ and the Binder cumulant $U_L$.
However, the distribution of $S$ is considerably broader, indicating
that the GAN has not managed to fully capture the dynamics. This
difference is ironed out by our statistical selection procedure, as we
will discuss next.

\subsection{Statistical Tests}

Following our arguments of \Cref{sec:algorithm}, we initially verified
the GAN's ability to reproduce every desired action value from field
configurations which were not part of the training dataset. Here, we
used $\varDelta S_{\mathrm{thresh}} = 1$ for the pre-sampling step and
a learning rate of $\epsilon = 10^{-5}$ for the gradient flow. In the
subsequent simulation runs, the GAN was always able to produce samples
with matching actions, suggesting that the algorithm can indeed be
sufficiently ergodic. In order to verify that the change of $P(z)$
under the gradient flow \Cref{eq:gradient} is negligible, we estimated
the squared distance $(z - z')^2$ and also compared the first four raw
moments of $P(z)$ and $P(z')$. To this end, we collected $10^5$ latent
vectors respectively from before and after the gradient flow
evolution. The squared difference was estimated to be $(1.048 \pm
0.003) \times 10^{-11}$. The small result indicates that suitable $z'$
do always exist in the neighborhood of $z$. The first four raw moments
of both distributions are observed to be equivalent within the given
error bounds, and in fact both the moments and errors were found to
be equivalent for several orders of magnitude below the significant
digits.

\begin{figure}[t!]
	\centering
	\includegraphics[width=\linewidth]{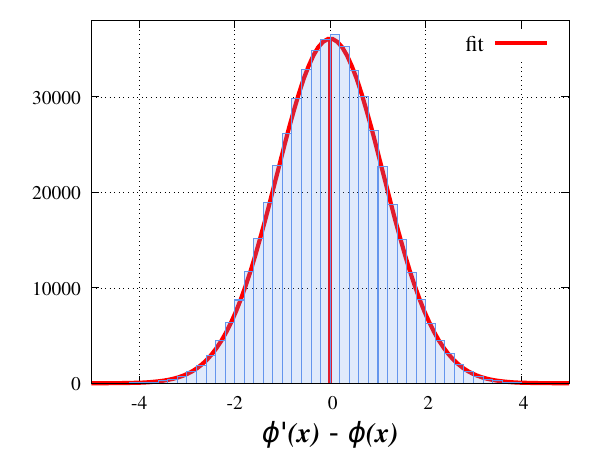}
	\caption{Histogram of differences of field variables before and after
	the GAN overrelaxation step, for a total of 500 steps, fitted with a
	Gaussian.}
	\label{fig:diffplot}
\end{figure}

\begin{table}[t]
	\centering
	\begin{tabular}{c || c | c }
		& $\langle M \rangle$ & \\
		\hline
		\rule{0pt}{3ex} training data: $10^3$ samples & 9.8e-04 $\pm$ 1.57e-03 & \\
		test data: $10^4$ samples & \ -8.87e-04 $\pm$ 4.99e-04 \ & \\
		$10^4$ generated samples & \ -7.18e-04 $\pm$ 5.01e-04 \ &
	\end{tabular}
	\ \\ \ \\ \ \\
	\begin{tabular}{c | c | c}
		& $\chi_2$ & $U_L$ \\
		\hline
		\rule{0pt}{3ex} & \ 2.52 $\pm$ 0.11 & \ -7.6e-03 $\pm$ 9.82e-02 \\
		& \ 2.55 $\pm$ 0.04 \ & \ 8.2e-03 $\pm$ 2.99e-02 \ \\ 		
		& \ 2.57 $\pm$ 0.04 \ & \ 4.9e-03 $\pm$ 2.99e-02 \ 
	\end{tabular}
	\caption{Comparison of observables calculated on different datasets
	with lower error bounds determined using the statistical jackknife
	method.}
	\label{tab:observables}
\end{table}

\Cref{fig:samples} shows a sample from the HMC simulation and three
corresponding example proposals generated with our overrelaxation
method. In \Cref{fig:diffplot} we plot a distribution of the
differences between individual field variables $\phi'(x) - \phi(x)$
before and after the overrelaxation for a total of 1000 independent
steps. By fitting a Gaussian to the histogram, we verified that the
distribution is symmetric and its mean consistent with zero.
Altogether, these numerical tests conclusively demonstrate the
validity of the given statistical arguments, suggesting that the
selection probability $T_0(\phi'|\phi)$ is indeed symmetric.
Furthermore, we tested and compared a variety of different latent
space dimensions $d_z$, starting at $N^2$ and successively going to
smaller values by powers of 2. The performances were consistent until
significant deviations were observed starting at $d_z = 32$, and the
GAN failed to converge entirely for $d_z = 8$ and below. This behavior
corresponds to the aforementioned plateau above a threshold value and
further supports the previous discussion of ergodicity in
\Cref{sec:algorithm}. For the results that follow, we chose $d_z =
256$.

\subsection{Efficiency Gain and Computational Cost}

\begin{figure*}
	\centering
	\includegraphics[width=0.49\linewidth]{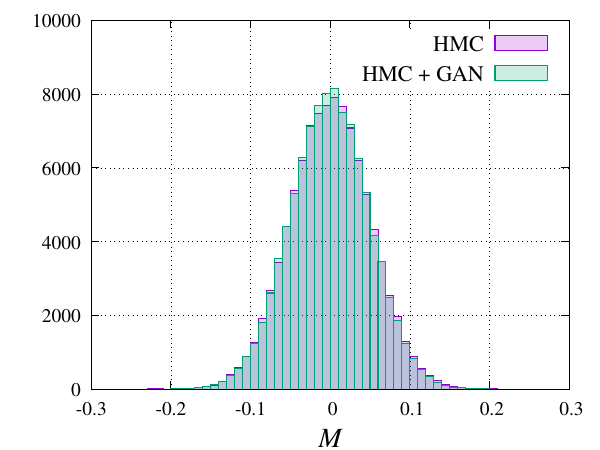}
	\includegraphics[width=0.49\linewidth]{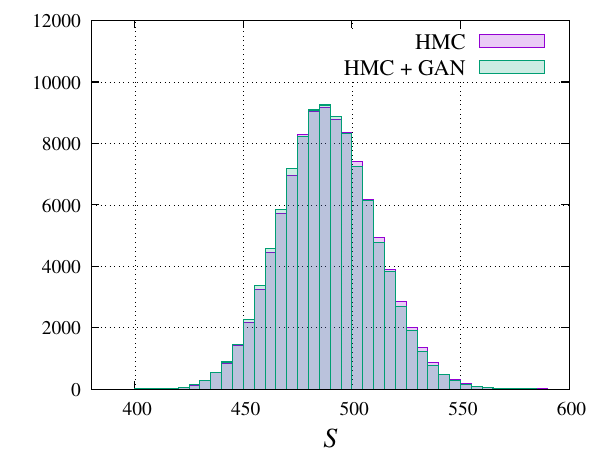}
	\caption{Comparison of magnetization (left) and action (right)
		distributions with $10^5$ samples generated with the baseline HMC and
		in combination with the GAN overrelaxation step using $n_H = 3$.}
	\label{fig:gan_overrelaxation}
\end{figure*}

\begin{figure}[t]
	\includegraphics[width=\linewidth]{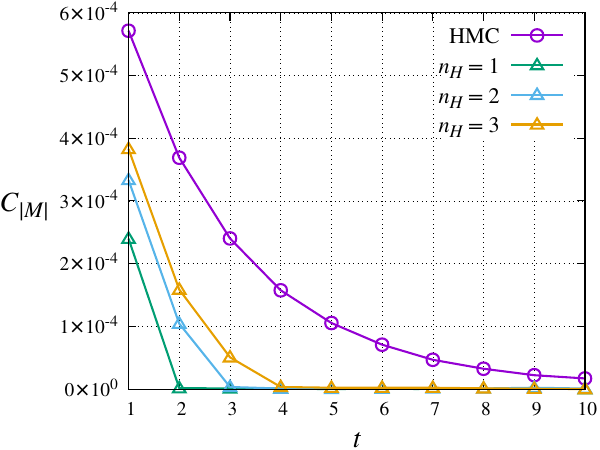}
	\caption{Comparison of the autocorrelation functions of $|M|$ for a
	local Metropolis update, the HMC algorithm and our method using $n_H
	\in \{1, 2, 3\}$. Results for $M$ also show the same qualitative
	behavior.}
	\label{fig:autocorr}
\end{figure}

The distributions of $M$ and $S$ obtained with our modified sampling
algorithm, as well as the aforementioned observables, are consistent
with results from the pure HMC simulation, see
\Cref{fig:gan_overrelaxation}. In particular, the difference between
the action distributions that was seen in \Cref{fig:gan_baseline} has
disappeared completely, which indicates that the algorithm correctly
reproduces the dynamics of the theory. We can estimate the efficiency
gain by comparing the behavior of the
associated autocorrelation functions $C_{|M|}(t)$, shown in
\Cref{fig:autocorr} for $n_H = 1,2,3$. We observe a substantial
reduction of autocorrelations by our method compared to the HMC
baseline, to the extent that $C_{|M|}(t)$ is almost zero at $t = n_H +
1$, i.e.\ after every GAN overrelaxation step. The small residual
autocorrelation observed for $t \geq n_H + 1$ stems solely from the
acceptance rate of the intermediate HMC steps and is exactly zero when
only accepted samples are taken into account for the computation of
$C_{|M|}(t)$.

We now determine for each Markov chain the corresponding integrated
autocorrelation time, defined as

\begin{align}
	\tau_{X,\text{int}} = \frac{1}{2} + \frac{1}{C_X(0)} \sum_{t=1}^{T} C_X(t)\ .
\end{align}
It is generally expected to scale as $\tau_{X,\text{int}} \sim
(\xi_X)^z$, where $\xi_X$ is now the correlation length for the
observable $X$ and $z$ again denotes the dynamical critical exponent,
which depends on the algorithm. We use $10^6$ consecutive measurements
of $|M|$ to calculate each $C_{|M|}(t)$ and truncate the sum for
$\tau_{|M|,\text{int}}$ at $T = 100$. For the HMC baseline, we obtain
$\tau_{|M|,\text{int}} \approx 2.29$. Using our modified algorithm,
the results for $n_H = 1,2,3$ are determined as $\tau_{|M|,\text{int}}
\approx 0.75,\,0.96,\,1.16$, respectively. This clearly demonstrates
that our proposed method could significantly improve the dynamical
critical exponents of established sampling algorithms.

\begin{figure}[t]
	\centering
	\includegraphics[width=\linewidth]{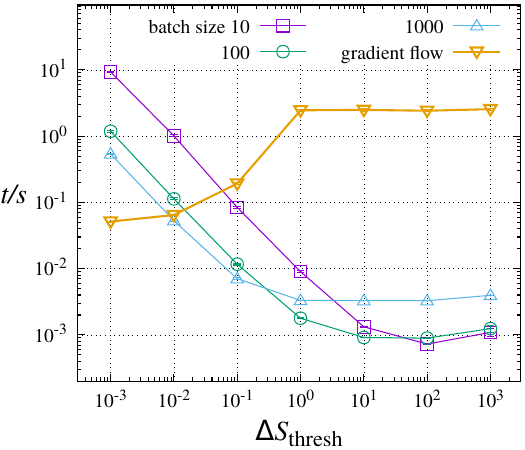}
	\caption{Average time given in seconds as a function of $\varDelta
	S_\text{thresh}$, both for the pre-sampling step using three
	different batch sizes, as well as the gradient flow step with a
	learning rate of $\epsilon = 10^{-6}$.}
	\label{fig:benchmark}
\end{figure}

In order to facilitate a rough comparison of the computational cost,
both the HMC update and the GAN overrelaxation step were implemented
using the same framework, the deep learning library \textsc{PyTorch}
\cite{PyTorch}. All steps of the simulation except for the recording
and monitoring functionality are performed on an Nvidia GeForce GTX
1070. The average time for the computation of one (accepted) HMC
trajectory was measured to be 42\,ms. For the GAN overrelaxation, the
average time depends on the aforementioned hyperparameters. In the
sampling step, we need to consider the behavior with respect to the
batch size and $\varDelta S_{\text{thresh}}$. Larger batches require
more time, but are more likely to contain suitable samples for small
values of $\varDelta S_{\text{thresh}}$ and are preferred in this
region, since repeatedly sampling smaller batches is less efficient.
On the contrary, if $\varDelta S_{\text{thresh}}$ is fixed at a larger
value, it is increasingly likely for any given sample to satisfy the
criterion, and small batch sizes are sufficient. \Cref{fig:benchmark}
shows the average time for three different batch sizes as a function
of $\varDelta S_{\text{thresh}}$. Plateaus can be observed at larger
and consistent scaling properties at smaller values.

In contrast to the time required for sampling, the gradient flow step
is completed faster for smaller $\varDelta S_{\text{thresh}}$ and
takes longer at large values, as shown in \Cref{fig:benchmark}. Hence,
one needs to find a trade-off value where neither of the two steps
takes a prohibitively long time. The gradient flow also generally
depends on the discrete step size or learning rate $\epsilon$, but
different choices of this parameter were in our case found to have
only a weak effect on the overall time. The optimal order of magnitude
for the hyperparameters was determined to be $\varDelta
S_{\text{thresh}} = 10^{-2}$ with a batch size of $10^3$ and a
learning rate of $\epsilon = 10^{-6}$. With these values, the sampling
and gradient flow step require on average 53\,ms and 64\,ms,
respectively, yielding a combined time of 117\,ms. Some modifications
which could further improve the performance are discussed in
\Cref{sec:concl}.

\section{Conclusions and Outlook}
\label{sec:concl}

In this work, we have investigated GANs for the purpose of reducing
autocorrelations in lattice simulations by providing independent field
configurations. A simple algorithm was constructed to enable a
statistically consistent injection of generated samples into a Markov
chain. We provide several numerical criteria to verify the validity of
this approach and demonstrate that it can improve standard importance
sampling algorithms.

To this end, we have implemented GANs as an overrelaxation step, which
can be achieved through an evolution of the latent variables. Putting
our reasoning to the test, we first confirmed that GANs can
successfully capture the dynamics of two-dimensional scalar
$\phi^4$-theory in the symmetric phase. We note here that, even though
we tested our ansatz using only a simple `vanilla' GAN, certain
properties were reproduced by it to remarkably high precision without
even employing a statistical selection procedure
(\Cref{fig:gan_baseline}, \Cref{tab:observables}). Deviations from the
action distribution were then eliminated completely
(\Cref{fig:gan_overrelaxation}). Furthermore, a significant reduction
of autocorrelations relative to the HMC baseline has been demonstrated
(\Cref{fig:autocorr}). A comparison of the integrated autocorrelation
times suggests that the incorporation of generative neural samplers
into lattice simulations could facilitate a considerable reduction of
dynamical critical exponents, which should be investigated in more
detail. We also emphasize here again that our method is compatible
with any action-based importance sampling algorithm and the observed
relative performance gain is expected to carry over.

The time benchmarks and hyperparameter optimization
(\Cref{fig:benchmark}) suggest that the computational cost of the
method could be further reduced by rather straightforward
modifications to the pre-sampling step. Most configurations are
immediately discarded until one with $|\varDelta S| < \varDelta
S_\text{thresh}$ is found. The repeated calculation of $S$ seems like
an unnecessary bottleneck. It may therefore be useful to store an
ensemble of field configurations from the GAN, together with their
associated actions, before starting the simulation. The reservoir
should then be repopulated periodically. This would ensure that
appropriate samples are readily available for the gradient flow step
and do not constantly need to be created and destroyed. With a
conditional GAN \cite{1411.1784} it may also be possible to reduce
$\varDelta S$ a priori by using the target value of the action as the
conditional parameter in the optimization. Training a conditional GAN
with samples generated at various different action parameters may also
allow extrapolation to otherwise numerically inaccessible regions of
the phase diagram where no training data is available in the first
place.

In conclusion, the findings presented here support generative neural
samplers as a method to accelerate lattice simulations. Our work
provides a comprehensive example of an improved sampling algorithm and
extensive numerical tests. First results regarding the reduction of
autocorrelation times are promising and encourage further research
into the matter.

\section*{Acknowledgements}
We thank L.~Kades, J.~Massa, M.~M\"uller, M.~Scherzer,
I.-O.~Stamatescu, N.~Strodthoff, S.~J.~Wetzel and F.~P.G.~Ziegler for
discussions. This work is supported by the Deutsche
Forschungsgemeinschaft (DFG, German Research Foundation) under
Germany's Excellence Strategy EXC 2181/1 - 390900948 (the Heidelberg
STRUCTURES Excellence Cluster) and under the Collaborative Research
Centre SFB 1225 (ISOQUANT), EMMI and the BMBF grant 05P18VHFCA.

\appendix
\bibliography{literature}{}
\bibliographystyle{utphys}

\end{document}